\documentclass[twocolumn,referee,epjc3]{svjour3} 
\smartqed  
\usepackage{float}
\usepackage{color,hyperref}
\usepackage{longtable}
\hypersetup{colorlinks,breaklinks,linkcolor=blue,urlcolor=blue,anchorcolor=blue,citecolor=blue}
\usepackage{amsmath}
\usepackage{mathrsfs}
\usepackage{mathtools} 
\usepackage{footmisc} 

\RequirePackage{graphicx}

\journalname{Eur. Phys. J. A}

\begin{document}

\title {Multiple chiral bands in $^{137}$Nd}

\author{C. M. Petrache\thanksref{1} \thanks {corresponding author: costel.petrache@csnsm.in2p3.fr} \and B. F. Lv\thanksref{2} \thanks {corresponding author: lvbingfeng@impcas.ac.cn}  \and  Q. B. Chen\thanksref{3} \and J. Meng\thanksref{4,5} \and A. Astier\thanksref{1} \and E. Dupont\thanksref{1} \and K. K. Zheng\thanksref{1} \and P. T. Greenlees\thanksref{6} \and H. Badran\thanksref{6} \and T. Calverley\thanksref{6,7} \and D. M. Cox\thanksref{6,8} \and T. Grahn\thanksref{6} \and J. Hilton\thanksref{6,7} \and R. Julin\thanksref{6} \and S. Juutinen\thanksref{6} \and J. Konki\thanksref{6,9} \and J. Pakarinen\thanksref{6} \and P. Papadakis\thanksref{6,10} \and J. Partanen\thanksref{6} \and P. Rahkila\thanksref{6}  \and P. Ruotsalainen\thanksref{6} \and M. Sandzelius\thanksref{6} \and J. Saren\thanksref{6} \and C. Scholey\thanksref{6} \and J. Sorri\thanksref{6,11} \and S. Stolze\thanksref{6,12} \and J. Uusitalo\thanksref{6} \and B. Cederwall\thanksref{13} \and A. Ertoprak\thanksref{13} \and H. Liu\thanksref{13} \and S. Guo\thanksref{2} \and J. G. Wang\thanksref{2} \and X. H. Zhou\thanksref{2} \and I. Kuti\thanksref{14} \and J. Tim\'ar\thanksref{14} \and A. Tucholski\thanksref{15} \and J. Srebrny\thanksref{15} \and C. Andreoiu\thanksref{16}}

\institute{Centre de Sciences Nucl\'eaires et Sciences de la  Mati\`ere, CNRS/IN2P3, Universit\'{e} Paris-Saclay, B\^at. 104-108, 91405  Orsay, France \label{1}
\and Institute of Modern Physics, Chinese Academy of Sciences, Lanzhou 730000, China \label{2}
\and Physik-Department, Technische Universit\"at M\"unchen, D-85747 Garching, Germany\label{3}
\and State Key Laboratory of Nuclear Physics and Technology, School of Physics, Peking University, Beijing 100871, China\label{4}
\and Yukawa Institute for Theoretical Physics, Kyoto University, Kyoto 606-8502, Japan \label{5}
\and Department of Physics, University of Jyv\"askyl\"a, FI-40014 Jyv\"askyl\"a, Finland  \label{6}
\and Department of Physics, University of Liverpool, The Oliver Lodge Laboratory, Liverpool L69 7ZE, United Kingdom \label{7}
\and Department of Mathematical Physics, Lund Institute of Technology, S-22362 Lund, Sweden \label{8}
\and CERN, CH-1211 Geneva 23, Switzerland \label{9}
\and STFC Daresbury Laboratory, Daresbury, Warrington, WA4 4AD, UK \label{10}
\and Sodankyl\"a Geophysical Observatory, University of Oulu, FIN-99600 Sodankyl\"a, Finland \label{11}
\and Physics Division, Argonne National Laboratory, Argonne, Illinois 60439, USA \label{12}
\and KTH Department of Physics,S-10691 Stockholm, Sweden \label{13}
\and Institute for Nuclear Research, Hungarian Academy of Sciences, Pf. 51, 4001 Debrecen, Hungary \label{14}
\and University of Warsaw, Heavy Ion Laboratory, Pasteura 5a, 02-093 Warsaw, Poland \label{15}
\and Department of Chemistry, Simon Fraser University, Burnaby, BC V5A 1S6, Canada \label{16}}
\renewcommand{\thefootnote}{\fnsymbol{footnote}}

\maketitle 

\abstract
{Two new bands have been identified in $^{137}$Nd from a high-statistics JUROGAM II gamma-ray spectroscopy experiment. Constrained density functional theory and particle rotor model calculations are used to assign configurations and investigate the band properties, which are well described and understood. It is demonstrated that these two new bands can be interpreted as chiral partners of previously known three-quasiparticle positive- and negative-parity bands. The newly observed chiral doublet bands in $^{137}$Nd represent an important support to the existence of multiple chiral bands in nuclei. The present results constitute the missing stone in the series of Nd nuclei showing multiple chiral bands, which becomes the most extended sequence of nuclei presenting multiple chiral bands in the Segr\'e chart.}


\section{Introduction}
The nuclei of the $A\approx130$ mass region constitute the largest ensemble of chiral nuclei \cite{Stefan-Jie} in the chart of the nuclides \cite{adndt-chiral}. Reviews of the experimental results and their theoretical interpretation can be found in Refs. \cite{frauendorf-2018,raduta-2016,meng-2016,meng-2014,Meng2010JPhysG37}.
Bands built on configurations involving three-, four- and six-quasiparticles have been observed in this mass region ($^{133}$La \cite{133la-petrache}, $^{133}$Ce \cite{Ayangeakaa2013Phys.Rev.Lett.172504}, $^{135}$Nd \cite{135nd-zhu,135nd-lv}, $^{136}$Nd \cite{136nd-PRCR,136nd-qibo} and $^{138}$Nd \cite{138nd-tilted,raduta-jpg-2016,raduta-jpg-2017}).
The present work is devoted to the study of chirality in $^{137}$Nd, the odd-even neighbor of $^{135}$Nd and $^{136}$Nd, in which multiple chiral doublet (M$\chi$D) \cite{Meng2006Phys.Rev.C037303} bands have been recently identified \cite{135nd-lv,136nd-PRCR,136nd-qibo}. Prior to this work, the $^{137}$Nd nucleus has been investigated both experimentally \cite{137nd-plb,137nd-decay,137nd-dipole,137nd-npa,137nd-oblate} and theoretically \cite{137nd-brant}.  Two new dipole bands are identified at medium spins, which are interpreted as chiral partners of previously known three-quasiparticle positive- and negative-parity bands. The interpretation of the three-quasiparticle bands and the chirality, previously investigated using the interacting boson model plus broken pairs \cite{137nd-brant}, are revisited. The resemblance between the new negative-parity chiral doublet of $^{137}$Nd with that known in $^{135}$Nd and $^{133}$Ce, gives strong support to the interpretation in terms of chiral vibration at low spin, and to the invoked transition between chiral vibration and chiral rotation in the odd-even Nd nuclei \cite{chiral-vibration}. The existence of two chiral doublet bands in $^{137}$Nd also gives a strong support to the existence of the M$\chi$D phenomenon in the $A\approx130$ mass region, which presents, in addition to $^{133}$Ce, in the most extended sequence of nuclei including both odd-even and even-even nuclei, from $^{135}$Nd to $^{138}$Nd. 
 The band structure is discussed within the constrained covariant density functional theory (CDFT) framework \cite{Meng2006Phys.Rev.C037303,meng2016} and the particle rotor model (PRM) recently developed to include multi-$j$ configurations, which is a powerful tool in the investigation of the 3D chiral geometry in nuclei \cite{136nd-qibo,135nd-prm,streck2018}. The present experimental results are obtained from the same data set from which were obtained the results of Refs.  \cite{135nd-lv,136nd-PRCR,137nd-oblate,lv,136nd-hd}. 

\section{Experimental results}
High-spin states in $^{137}$Nd were populated using the $^{100}$Mo($^{40}$Ar,3n) reaction at a  beam energy of 152 MeV, provided at the Accelerator Laboratory of the University of Jyv\"askyl\"a, Finland. A self-supporting enriched $^{100}$Mo foil of 0.50 mg/cm$^{2}$ thickness was used as a target.
The JUROGAM II array \cite{JUROGAM} consisting of 24 clover and 15 coaxial tapered germanium detectors placed at  the target position was used to detect prompt $\gamma$-rays.  A total of $5.1\times10^{10}$ prompt $\gamma$-ray coincidence events with fold $\geq$ 3 were collected.  All the data were recorded by the triggerless Total Data Readout (TDR) data acquisition \cite{lazarus} and the events were time-stamped using a 100 MHz clock. The data were sorted using the GRAIN code \cite{grain}. Fully symmetrized, three-dimensional ($E_{\gamma}$-$E_{\gamma}$-$E_{\gamma}$) and four-dimensional ($E_{\gamma}$-$E_{\gamma}$-$E_{\gamma}$-$E_{\gamma}$) matrices were analyzed using the \textsc{radware}~\cite{rad1,rad2} analysis package.  

\begin{figure*}[ht]
\vskip -.cm
\rotatebox{-90}{\scalebox{0.63}{\includegraphics[width=\textwidth]{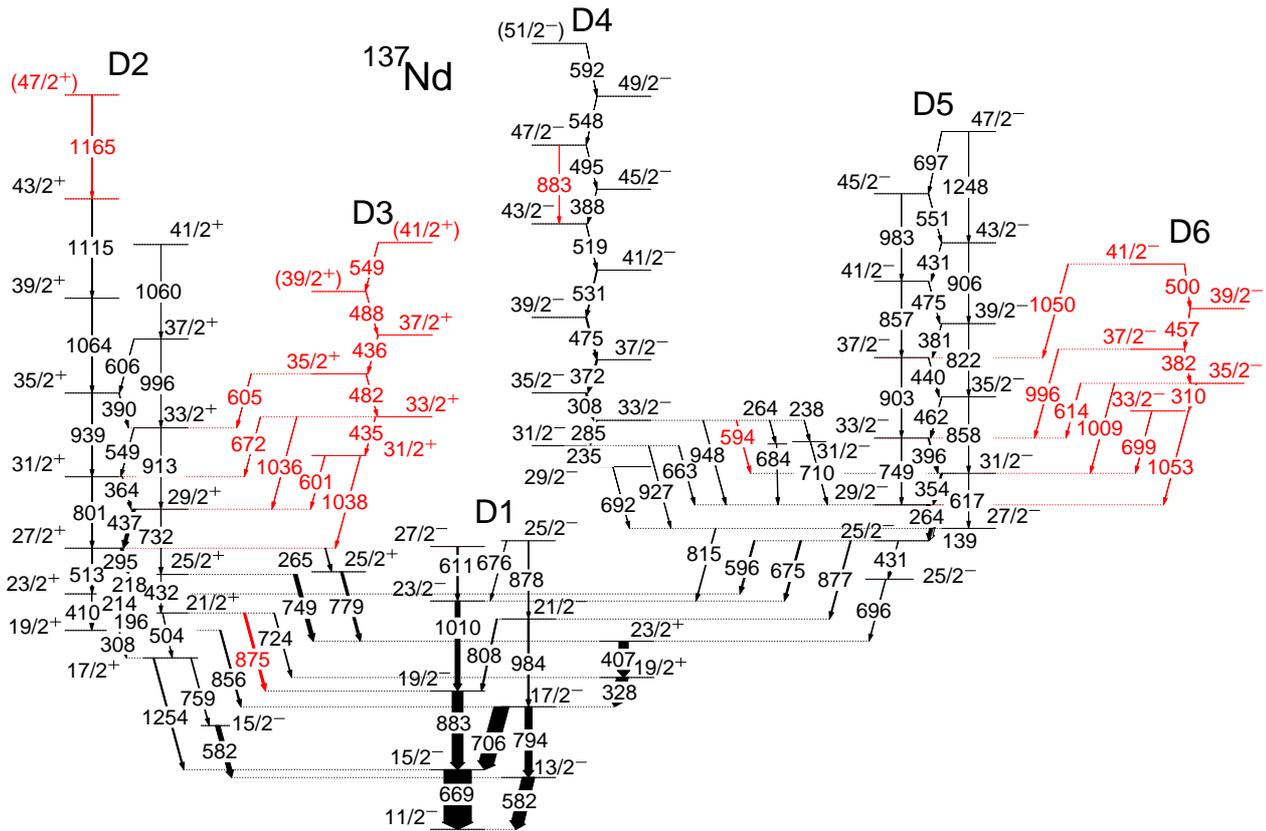}}}
\vskip .cm
\caption []{(Color online) Partial level scheme of  $^{137}$Nd showing, in addition to the previously known bands D1, D2 and D5,  the newly identified bands D3 and D6, and 594-, 875-, 883- and 1165-keV transitions (marked in red). }
\label{fig1} 
\end{figure*}

The partial level scheme of $^{137}$Nd showing in red the newly identified bands is given in Fig. \ref{fig1}.
Spin and parity assignments for newly observed levels are based on the measured two-point angular correlation (anisotropy) ratios R$_{ac}$ \cite{Rac0,Rac1}.
To extract the  R$_{ac}$ values, the data were sorted into $\gamma$-$\gamma$ matrices constructed by sorting prompt coincidence events with the detectors mounted at ($133.6^{\circ}$ and $157.6^{\circ}$) versus (all angles) and ($75.5^{\circ}$ and $104.5^{\circ}$) versus (all angles) combinations, by setting the same energy gates on the (all angles) projection spectrum in both matrices, and projecting on the other axis. Then, the R$_{ac}$ ratio was calculated using the extracted relative intensities of the $\gamma$-rays of interest ($I_{\gamma}$) from these spectra, normalized by the different efficiencies of the two sets of detectors. To determine the efficiency at different angles, we combined all tapered germanium detectors mounted at $133.6^{\circ}$ and  $157.6^{\circ}$ in one ring, and  all clover detectors mounted around 90$^\circ$ ($75.5^{\circ}$ and $104.5^{\circ}$) in one ring, respectively. 
The R$_{ac}$ values for stretched dipole and quadrupole transitions are around 0.8 and 1.4, respectively, have been deduced from the analysis of strong E2, E1, and M1 transitions in $^{`36}$Nd. More details of the experimental setup and data analysis can be also found in Ref. \cite{lv}. The $\gamma$-ray energies (E$_\gamma$), level energies (E$_i$), relative intensities (I$_\gamma$), R$_{ac}$ values, multipolarities, and assigned spin-parity of the different $\gamma$-ray transitions of $^{137}$Nd are listed in Table \ref{table1}.

\begin{figure*}[ht]
\centering
\hskip .cm
\vskip 3.cm
\rotatebox{-0}{\scalebox{0.68}{\includegraphics[width=\textwidth]{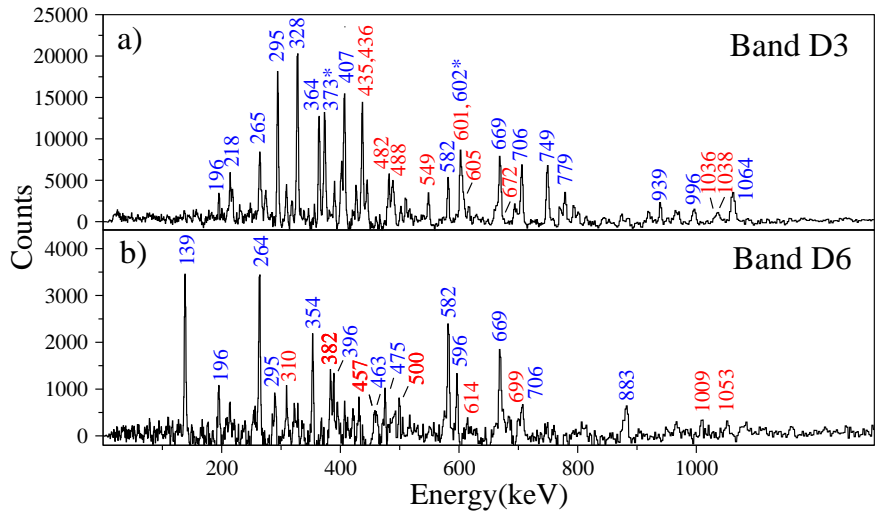}}}
\caption []{(Color online) Gamma-ray spectra for bands D3 and D6 of $^{137}$Nd obtained by double-gating on selected in- and out-of-band transitions: 436, 482, 488, 1036, 1038 keV for band D3, and 310, 382, 457, 996, 1009, 1053 keV for band D6. Newly identified transitions are indicated in red color. The contaminant transitions of 373 and 602 keV belonging to $^{136}$Nd populated by the strongest reaction channel are indicated with an asterisk.}
\label{fig2} 
\end{figure*}

\begin{figure*}[ht]
\centering
\hskip .cm
\vskip 3.cm
\rotatebox{-0}{\scalebox{0.79}{\includegraphics[width=\textwidth]{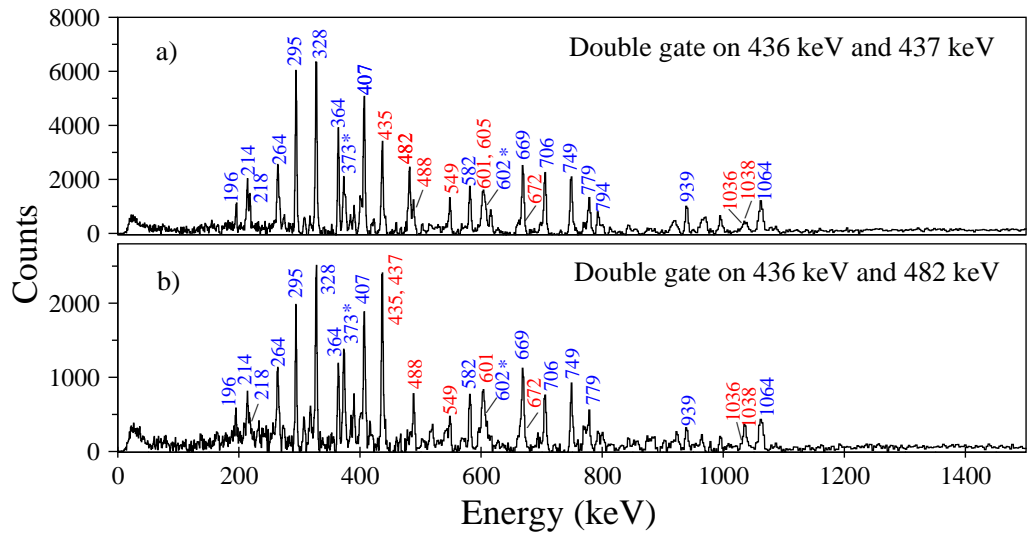}}}
\vskip -0.cm
\caption []{(Color online) Gamma-ray spectra for band D3 of $^{137}$Nd obtained by double-gating on 436 keV and 437 keV in panel a), and on 436 keV and 482 keV in panel b). Newly identified transitions are indicated in red color. The contaminant transitions of 373 and 602 keV belonging to $^{136}$Nd populated by the strongest reaction channel are indicated with an asterisk.}
\label{fig3} 
\end{figure*}

\begin{figure*}[ht]
\centering
\hskip .cm
\vskip 3.cm
\rotatebox{-0}{\scalebox{0.79}{\includegraphics[width=\textwidth]{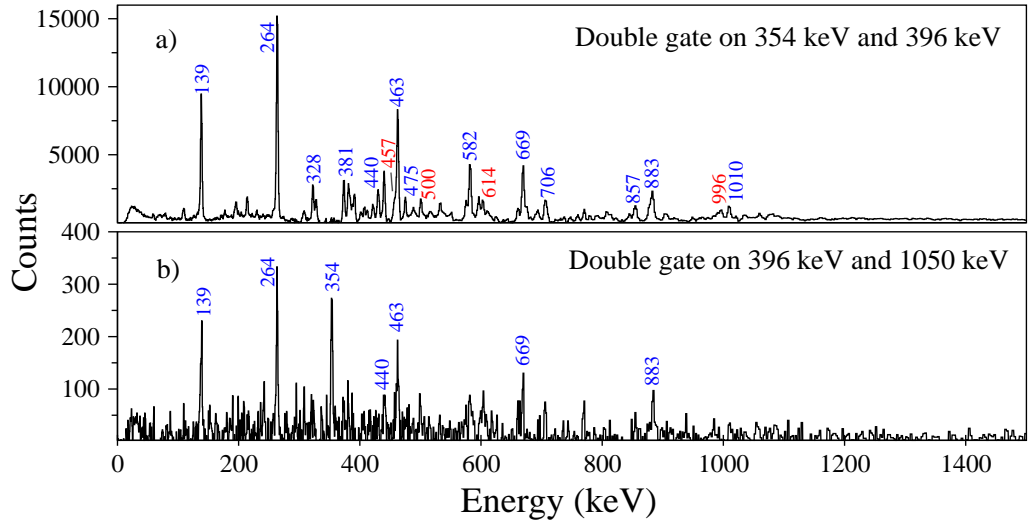}}}
\vskip -0.cm
\caption []{(Color online)  Gamma-ray spectra for the 996- and 1050-keV transitions connecting bands D5 and D6 of $^{137}$Nd, obtained by double-gating on 354 keV and 396 keV in panel a), and on 396 keV and 1050 keV in panel b). Newly identified transitions are indicated in red color.}
\label{fig4} 
\end{figure*}

\clearpage
\onecolumn
\begin{center}
\begin{longtable}{cccccc}
\caption{\label{table1}Experimental information including the $\gamma$-ray energies, energies of the initial levels E$_{i}$, intensities I$_\gamma$, R$_{ac}$, multipolarities, and the spin-parity assignments to the observed states in $^{137}$Nd.}\\
\hline{}
$\gamma$-ray     energy $^a$ & E$_i$ (keV)   & Intensity $^b$ & R$_{ac}$$^c$ & Multipolarity & J$^{\pi}_i$ $\rightarrow$ J$^{\pi}_f$\\
\hline
\endfirsthead
\multicolumn{6}{c}
{\tablename\ \thetable\ -- \textit{Continued}} \\
\hline{}
$\gamma$-ray     energy $^a$ & E$_i$ (keV)   & Intensity $^b$ & R$_{ac}$$^c$ & Multipolarity & J$^{\pi}_i$ $\rightarrow$ J$^{\pi}_f$\\
\hline
\endhead
\hline \multicolumn{6}{r}{\textit{}} \\
\endfoot
\hline
\endlastfoot

$\bf  Band ~D1$ & & &         &            &                                                \\
          581.8               & 1100.8         &55(4)                       &0.82(3)                & M1/E2      & 13/2$^-$ $\rightarrow$ 11/2$^-$  \\
          610.6               & 3691.6         &4.9(6)                      & 1.4(1)                 & E2           & 27/2$^-$ $\rightarrow$ 23/2$^-$  \\
          675.9               & 3756.9         &0.13(5)                    &                           &                 & 25/2$^-$ $\rightarrow$ 23/2$^-$  \\
          669.4               & 1188.4         &100                          &1.38(8)                & E2           & 15/2$^-$ $\rightarrow$ 11/2$^-$  \\
          706.2               & 1894.8         &54(4)                       &0.74(7)                & M1/E2     & 17/2$^-$ $\rightarrow$ 15/2$^-$  \\
          793.8               & 1894.8         &25.8(31)                  &1.42(9)                & E2           & 17/2$^-$ $\rightarrow$ 13/2$^-$  \\   
          807.8               & 2879.1         &4.4(4)                      &1.1(2)                  & M1/E2     & 21/2$^-$ $\rightarrow$ 19/2$^-$  \\
          877.8               & 3756.9         &0.99(31)                  & 1.41(18)             & E2           & 25/2$^-$ $\rightarrow$ 21/2$^-$  \\
          882.9               & 2071.3         &40(2)                       &1.44(6)                & E2           & 19/2$^-$ $\rightarrow$ 15/2$^-$  \\          
          984.5               & 2879.1         &5.8(5)                      &1.51(12)              & E2            & 21/2$^-$ $\rightarrow$ 17/2$^-$  \\
         1009.7               & 3081.0        &18.1(7)                    &1.36(5)                & E2           & 23/2$^-$ $\rightarrow$ 19/2$^-$  \\

  $\bf  Band ~D2$ & & &         &            &                                                \\
          195.9               & 2946.6         & 3.1(2)                          &0.77(5)                      & M1/E2               & 21/2$^+$ $\rightarrow$ 19/2$^+$  \\  
          214.0               & 3160.6         & 7.7(8)                          & 0.79(7)                     & M1/E2               & 23/2$^+$ $\rightarrow$ 21/2$^+$  \\
          218.1               & 3378.7         &1.87(38)                       & 0.5(2)                       & M1/E2         & 25/2$^+$ $\rightarrow$ 23/2$^+$  \\
          265.0                & 3673.5        & 2.7(4)                          &1.1(1)                        & M1/E2         & 27/2$^+$ $\rightarrow$ 25/2$^+$  \\
          294.8               & 3673.5         & 9(1)                             & 0.81(4)                     & M1/E2              & 27/2$^+$ $\rightarrow$ 25/2$^+$  \\
          308.0                & 2750.7        &2.8(3)                           &0.79(3)                      & M1/E2              & 19/2$^+$ $\rightarrow$ 17/2$^+$  \\
          364.1               & 4474.4         & 2.01(24)                      & 0.88(9)                     & M1/E2              & 31/2$^+$ $\rightarrow$ 29/2$^+$    \\
          390.3               & 5413.5         & 0.47(23)                      &                                 &                    & 35/2$^+$ $\rightarrow$ 33/2$^+$  \\
          409.9               & 3160.6         & 1.67(48)                      & 1.54(31)                   & E2              & 23/2$^+$ $\rightarrow$ 19/2$^+$  \\
          432.1               & 3378.7         & 0.34(16)                      &                                 &                    & 25/2$^+$ $\rightarrow$ 21/2$^+$  \\
          436.8               & 4110.3         & 11.7(27)                       & 0.8(1)                      & M1/E2              & 29/2$^+$ $\rightarrow$ 27/2$^+$  \\
          503.9               & 2946.6         & 1.1(1)                          & 1.47(9)                     & E2              & 21/2$^+$ $\rightarrow$ 17/2$^+$  \\
          512.9               & 3673.5         & 0.78(39)                      &                                 &               & 27/2$^+$ $\rightarrow$ 23/2$^+$  \\
          548.8               & 5023.2         & 0.93(16)                      & 0.71(17)                  & M1/E2              & 33/2$^+$ $\rightarrow$ 31/2$^+$  \\
          582.5                & 1683.3        &16.4(22)                       &0.65(14)                   & M1/E2         & 15/2$^-$ $\rightarrow$ 13/2$^-$  \\
          605.8               & 6019.3         & 0.70(27)                      & 0.8(3)                      & M1/E2              & 37/2$^+$ $\rightarrow$ 35/2$^+$    \\
          723.9                & 2946.6        & 2.7(2)                          & 1.21(14)                  & M1/E2         & 21/2$^+$ $\rightarrow$ 19/2$^+$  \\
          731.6               & 4110.3         & 1.80(62)                      & 1.5(2)                       &E2               & 29/2$^+$ $\rightarrow$ 25/2$^+$  \\
          749.3                & 3378.7        &13.9(15)                       &0.81(6)                     & M1/E2               & 25/2$^+$ $\rightarrow$ 23/2$^+$  \\         
          759.4                & 2442.7        & 2.6(4)                          & 0.83(5)                     & E1              & 17/2$^+$ $\rightarrow$ 15/2$^-$  \\     
          779.1                & 3408.5        & 8.7(5)                          &0.74(7)                      & M1/E2              & 25/2$^+$ $\rightarrow$ 23/2$^+$  \\    
          801.1               & 4474.4         &2.62(40)                       & 1.32(20)                   & E2              & 31/2$^+$ $\rightarrow$ 27/2$^+$    \\
          856.1                & 2750.7        &5.0(4)                           & 0.85(6)                     & E1              & 21/2$^+$ $\rightarrow$ 19/2$^-$  \\
          875.3                & 2946.6        & 8.1(6)                          & 0.78(9)                     & E1                 & 21/2$^+$ $\rightarrow$ 19/2$^-$  \\
          912.9               & 5023.2         & 0.58(20)                      &  1.41(19)                  & E2           & 33/2$^+$ $\rightarrow$ 29/2$^+$    \\
          939.1               & 5413.5         &1.6(2)                           & 1.35(15)                   & E2           & 35/2$^+$ $\rightarrow$ 31/2$^+$  \\
          996.1               & 6019.3         & 0.45(18)                      & 1.6(3)                          & E2           & 37/2$^+$ $\rightarrow$ 33/2$^+$  \\
         1060.2               & 7079.5        & 0.4(1)                          & 1.36(31)                     & E2           & 41/2$^+$ $\rightarrow$ 37/2$^+$    \\
         1063.8               & 6477.3        & 1.36(15)                      & 1.49(16)                     & E2           & 39/2$^+$ $\rightarrow$ 35/2$^+$    \\
         1115.2               & 7592.5        & 0.5(1)                          &1.46(13)                       & E2           & 43/2$^+$ $\rightarrow$ 39/2$^+$  \\
         1164.8               & 8757.3        & $<$0.1                        &                                   &                 & (47/2$^+)$$\rightarrow$ 43/2$^+$  \\  
         1254.3              & 2442.7         & 4.7(3)                          & 0.77(8)                     & E1                 & 17/2$^+$ $\rightarrow$ 15/2$^-$  \\

 $\bf  Band ~D3 $          & &            &                 &                      &         \\
       435.0               & 5146.7         & 0.4(1)                       & 1.2(3)                          & M1/E2              & 33/2$^+$ $\rightarrow$ 31/2$^+$  \\
       436.3               & 6064.7         & 0.6(2)                       & 0.91(27)                      & M1/E2                   & 37/2$^+$ $\rightarrow$ 35/2$^+$  \\
       481.7               & 5628.4         & 0.5(1)                       & 0.82(14)                      & M1/E2                   & 35/2$^+$ $\rightarrow$ 33/2$^+$  \\
       487.8               & 6552.5         & 0.13(3)                     &                                    &                         & (39/2$^+)$ $\rightarrow$ 37/2$^+$  \\
       549.1               & 7101.6         & $<$0.1                     &                                    &                         & (41/2$^+)$ $\rightarrow$(39/2$^+)$  \\
       601.4               & 4711.7         & 0.42(15)                   &                                    &                         & 31/2$^+$ $\rightarrow$ 29/2$^+$  \\    
       605.2               & 5628.4         & 0.33(7)                     & 0.64(16)                     & M1/E2                   & 35/2$^+$ $\rightarrow$ 33/2$^+$   \\
       672.3               & 5146.7         & 0.2(1)                       &                                   &                         & 33/2$^+$ $\rightarrow$ 31/2$^+$  \\
       1036.4              & 5146.7         & 0.45(9)                    & 1.34(29)                     & E2                    & 33/2$^+$ $\rightarrow$ 29/2$^+$    \\ 
       1038.2              & 4711.7         & 0.22(7)                    & 1.47(40)                     & E2                    & 31/2$^+$ $\rightarrow$ 27/2$^+$    \\
       
  $\bf Band ~D4$ & &            &                 &                  &         \\  
         
          234.8               & 4822.2         & 0.7(1)                       & 0.83(5)                       & M1/E2                 & 31/2$^-$ $\rightarrow$ 29/2$^-$  \\ 
          238.5               & 5107.5         &0.81(7)                     & 0.6(1)                          &M1/E2            & 33/2$^-$ $\rightarrow$ 31/2$^-$  \\ 
          264.4               & 5107.5         &1.2(1)                       & 0.86(7)                        &M1/E2                  & 33/2$^-$ $\rightarrow$ 31/2$^-$  \\ 
          285.3               & 5107.5         & 1.71(8)                    & 1.04(10)                      & M1/E2           & 33/2$^-$ $\rightarrow$ 31/2$^-$  \\
          307.8               & 5415.3         & 5.8(2)                      & 0.77(6)                        & M1/E2                 & 35/2$^-$ $\rightarrow$ 33/2$^-$  \\
          372.0               & 5787.3         & 5.4(7)                      & 1.2(2)                          & M1/E2           & 37/2$^-$ $\rightarrow$ 35/2$^-$  \\
          388.1               & 7700.0         & 0.9(1)                      & 0.73(6)                        & M1/E2                 & 45/2$^-$ $\rightarrow$ 43/2$^-$    \\
          474.8               & 6262.1         & 4.5(3)                      & 0.80(5)                        & M1/E2                 & 39/2$^-$ $\rightarrow$ 37/2$^-$  \\
          494.8               & 8194.8         & 0.46(4)                    &0.91(7)                         & M1/E2                 & 47/2$^-$ $\rightarrow$ 45/2$^-$    \\
          519.0               & 7311.9         & 1.6(2)                      &0.85(4)                         & M1/E2                 & 43/2$^-$ $\rightarrow$ 41/2$^-$  \\
          530.8               & 6792.9         & 3.3(3)                      &0.79(8)                         & M1/E2                 & 41/2$^-$ $\rightarrow$ 39/2$^-$  \\
          548.0               & 8742.6         & 0.24(4)                    &0.99(13)                       & M1/E2            & 49/2$^-$ $\rightarrow$ 47/2$^-$  \\
          592.3               & 9335.1         &$<$0.1                     &                                     &                       & (51/2$^-)$ $\rightarrow$ 49/2$^-$  \\
          594.5               & 5107.5         &1.9(3)                      & 1.1(2)                            &M1/E2            & 33/2$^-$ $\rightarrow$ 31/2$^-$  \\ 
          662.8               & 4822.2         &1.07                        &1.16(11)                          & M1/E2           & 31/2$^-$ $\rightarrow$ 29/2$^-$  \\
          683.7               & 4843.1         &1.45(9)                   &1.08(8)                            &M1/E2            & 31/2$^-$ $\rightarrow$ 29/2$^-$  \\
          691.8               & 4587.4         &0.75(9)                   &0.92(13)                          & M1/E2                 & 29/2$^-$ $\rightarrow$ 27/2$^-$    \\ 
          709.6               & 4869.0         &1.1(1)                     &0.74(9)                            &M1/E2                  & 31/2$^-$ $\rightarrow$ 29/2$^-$  \\ 
          882.9               & 8194.8         & 0.31(5)                  & 1.4(2)                             & E2                  & 47/2$^-$ $\rightarrow$ 43/2$^-$    \\      
          926.6               & 4822.2         &0.23(2)                   & 1.42(15)                         & E2                  & 31/2$^-$ $\rightarrow$ 27/2$^-$  \\
          948.1               & 5107.5         &0.88(6)                   &1.35(6)                            & E2                  & 33/2$^-$ $\rightarrow$ 29/2$^-$    \\

$\bf  Band ~D5$               &                &                            &                       &                       &                                 \\
          138.7               & 3895.6         & 11.7(15)                   & 0.79(4)               & M1/E2                    & 27/2$^-$ $\rightarrow$ 25/2$^-$  \\
          263.8               & 4159.4         & 11.9(16)                   & 0.84(8)               & M1/E2                    & 29/2$^-$ $\rightarrow$ 27/2$^-$  \\
          353.6               & 4513.0         & 5.6(3)                      & 0.74(7)                & M1/E2                    & 31/2$^-$ $\rightarrow$ 29/2$^-$  \\
          381.4               & 6192.5         & 0.6(1)                      & 1.01(14)              & M1/E2               & 39/2$^-$ $\rightarrow$ 37/2$^-$  \\
          395.5               & 4908.5         & 3.2(2)                      & 0.76(4)                & M1/E2                    & 33/2$^-$ $\rightarrow$ 31/2$^-$  \\
          431.1               & 7098.7         & 0.14(3)                    &                            &                          & 43/2$^-$ $\rightarrow$ 41/2$^-$  \\
          431.4               & 3756.4         &0.7(4)                       &                            &                          & 25/2$^-$ $\rightarrow$ 25/2$^-$  \\
          440.1               & 5811.1         & 0.99(7)                    & 0.78(5)                & M1/E2                    & 37/2$^-$ $\rightarrow$ 35/2$^-$  \\
          462.5               & 5371.0         & 2.1(1)                      & 0.82(3)                & M1/E2                    & 35/2$^-$ $\rightarrow$ 33/2$^-$  \\
          475.1               & 6667.6         & 0.40(5)                    & 1.25(21)              & M1/E2               & 41/2$^-$ $\rightarrow$ 39/2$^-$    \\
          551.4               & 7650.1         & 0.10(1)                    & 0.83(20)              & M1/E2                    & 45/2$^-$ $\rightarrow$ 43/2$^-$    \\
          595.8               & 3756.4         &5.1(3)                       &0.79(4)                 & E1                    & 25/2$^-$ $\rightarrow$ 23/2$^+$  \\
          617.4               & 4513.0         & 0.4(1)                      & 1.51(11)              & E2                    & 31/2$^-$ $\rightarrow$ 27/2$^-$    \\
          675.4               & 3756.4         &5.7(5)                       &0.79(6)                 &M1/E2                     & 25/2$^-$ $\rightarrow$ 23/2$^-$  \\
          696.1               & 3325.5         &1.8(2)                       & 0.85(7)                & E1                    & 25/2$^-$ $\rightarrow$ 23/2$^+$  \\
          697.0               & 8347.1         & $<$0.1                    &                             &                         & 47/2$^-$ $\rightarrow$ 45/2$^-$  \\
          749.1               & 4908.5         & 0.5(1)                      & 1.37(17)              & E2                    & 33/2$^-$ $\rightarrow$ 29/2$^-$  \\
          814.6               & 3895.6         &2.5(4)                       & 1.6(2)                  & E2                    & 27/2$^-$ $\rightarrow$ 23/2$^-$  \\  
          821.5               & 6192.5         & 0.10(2)                    & 1.45(24)              & E2                    & 39/2$^-$ $\rightarrow$ 35/2$^-$    \\
          856.5               & 6667.6         & 0.17(2)                    & 1.32(13)              & E2                    & 41/2$^-$ $\rightarrow$ 37/2$^-$  \\
          858.0               & 5371.0         & 0.6(2)                      & 1.44(15)              & E2                    & 35/2$^-$ $\rightarrow$ 31/2$^-$  \\       
          877.3               & 3756.4         & 3.8(4)                      &1.43(9)                 & E2                   & 25/2$^-$ $\rightarrow$ 21/2$^-$  \\
          902.6               & 5811.1         & 0.15(3)                    & 1.59(28)               & E2                  & 37/2$^-$ $\rightarrow$ 33/2$^-$  \\
          906.2               & 7098.7         & 0.11(1)                    & 1.45(19)               & E2                  & 43/2$^-$ $\rightarrow$ 39/2$^-$  \\
          982.5               & 7650.1         & 0.05(1)                    &                              &                       & 45/2$^-$ $\rightarrow$ 41/2$^-$  \\
          1248.4             & 8347.1         & 0.10(1)                    & 1.61(34)               & E2                  & 47/2$^-$ $\rightarrow$ 43/2$^-$  \\

$\bf  Band ~D6$ & & &         &            &                                                \\
           310.0                & 5522.4         &0.19(5)                & 0.84(10)                  & M1/E2              & 35/2$^-$ $\rightarrow$ 33/2$^-$  \\
           382.2                & 5904.6         & 0.17(4)               & 0.77(12)                  & M1/E2              & 37/2$^-$ $\rightarrow$ 35/2$^-$  \\
           456.6                & 6361.2         & 0.12(4)               & 0.54(21)                  & M1/E2           & 39/2$^-$ $\rightarrow$ 37/2$^-$    \\
           500.2                & 6859.0         &0.11(2)                & 1.07(27)                  & M1/E2           & 41/2$^-$ $\rightarrow$ 39/2$^-$    \\
           613.9                & 5522.4         &0.30(9)                & 0.67(19)                  & M1/E2              & 35/2$^-$ $\rightarrow$ 33/2$^-$  \\ 
           699.4                & 5212.4         &0.17(4)                & 0.58(12)                  & M1/E2           & 33/2$^-$ $\rightarrow$ 31/2$^-$  \\
           996.1                & 5904.6         & 0.38(7)               & 1.45(23)                  & E2              & 37/2$^-$ $\rightarrow$ 33/2$^-$  \\
          1009.4                & 5522.4         &0.22(8)               & 1.4(2)                      & E2              & 35/2$^-$ $\rightarrow$ 31/2$^-$  \\
          1050.3                & 6861.4         &$\le$0.05             &                              & E2              & 41/2$^-$ $\rightarrow$ 37/2$^-$    \\
          1053.0                & 5212.4         &0.14(3)                & 1.43(27)                  & E2              & 33/2$^-$ $\rightarrow$ 29/2$^-$  \\

\end{longtable}
\end{center}
$^a$ The  error on the transition energies is 0.2 keV for transitions below 1000 keV of the $^{137}$Nd reaction channel, 
0.5 keV for transitions above 1000 keV, and 1 keV for transitions above 1200 keV. \\
$^b$ Relative intensities corrected for efficiency, normalized to the intensity of the 699.4-keV transition. The transition intensities were obtained from a combination of total projection and gated spectra.  \\
$^c$ The R$_{ac}$ has been deduced from two asymmetric $\gamma$-$\gamma$ coincidence 
matrices sorted with all detectors on one axis, and detectors around 90$^{\circ}$ and at backward
 angles, respectively, on the other axis. The tentative spin-parity of the states are given in parenthesis.\\
\clearpage
\twocolumn


 The bands D1, D2, D4 and D5 were previously reported in Ref. \cite{137nd-npa}. Their spins and parities were well established on the basis of directional correlation ratios from oriented states reported in Ref. \cite{137nd-npa}. In the present work, we mainly focus on the newly observed bands D3 and  D6. In addition, the following new transitions have been identified in the previously reported bands: 1164.8 keV on top of band D2, 875.3 keV connecting band D2 to band D1, 882.9 keV in band D4, and 594.5 keV connecting band D4 to band D5. 

The positive-parity band D3 consisting of six levels with spins from $31/2^+$ to $(41/2^+)$,  inter-connected by the 435-, 482- 436-, 488- and 549-keV dipole transitions, has been newly identified. It decays to band D2 by five transitions of 601, 605, 672, 1036 and 1038 keV, which are assigned M1 and E2 characters. The alternative E1 or M2 characters are less probable because the existence of several E1 transitions would imply either enhanced dipole moments which are not expected to be present in this nucleus, and the existence of several M2 transitions would imply large change of angular momentum and change of parity, which hardly can compete with collective E2 transitions at high spins.   Fig. \ref{fig2}(a) shows the newly identified in-band and out-of-band transitions of bands D3. It should be pointed out that the contaminant transitions of 373 and 602 keV belonging to $^{136}$Nd populated by the strongest reaction channel, are present in Fig. \ref{fig2}(a) due to the used sum of gates on transitions with energies close to those of transitions in $^{136}$Nd (481 keV of band D2, 485 keV of band D1, 486 keV of band N1, 487 keV of bands D4 and L3, 488 keV of $\gamma$-band,  1039 keV of band T2, all reported in Ref. \cite{lv}); however, the spectrum is dominated by the 196-, 218-, 295-, 328-, 407, 582-, 669- and 706-keV transitions  of $^{137}$Nd, giving clear evidence for the assignment of the new band structure to $^{137}$Nd. The existence of three transitions with energies around 436 keV, one in band D2 and two in band D3 is demonstrated in Fig. \ref{fig3}, where one can see the presence of the 435-keV peak is the double-gated spectrum on the 436- and 437-keV transitions.
The spins and parity of band D3 are assigned based on the R$_{ac}$ values of the 605-, 1036-, and 1038-keV connecting transitions to band D2, in particular those of 1038 and 1036 keV which have E2 character (see Table \ref{table1}). 

 The negative-parity band D6 consisting of five levels with spins from $33/2^-$ to $41/2^-$, inter-connected by the 310-, 382-, 457- and 500-keV dipole transitions, has been also newly identified. It decays to band D5 by six transitions of 614, 699, 996, 1009, 1050 and 1053 keV, which are assigned M1 and E2 character based on considerations similar to those of band D3.  
A double-gated spectrum showing the newly identified in-band and out-of-band transitions of band D6 is given in Fig. \ref{fig2}(b).
In order to identify the weak 996- and 1050-keV connecting transitions between the bands D6 and D5 not clearly seen in the spectrum of Fig. \ref{fig2}(b), double-gated spectra like those of Fig. \ref{fig4} were used. Based on the R$_{ac}$ values  of the 1053- and 699-keV decay-out transitions which clearly indicate their quadrupole and dipole nature, respectively  (see Table \ref{table1}), a spin 33/2 and negative parity have been assigned to the band-head of band D6. The R$_{ac}$ values  of three other inter-band transitions between bands D5 and D6 have also been measured, confirming the spin and parity assignment to the levels of band D6. 



\section{Discussion}

We investigated the structure of the bands discussed in the present work using the constrained CDFT and PRM. 
The unpaired nucleon configurations, quadrupole deformation parameters $(\beta, \gamma)$ obtained by CDFT with PC-PK1 interaction \cite{Zhao2010Phys.Rev.C054319} as well as the quadrupole deformation parameters $(\beta^\prime,\gamma^\prime)$, moments of inertia $\mathcal{J}_0$ (unit $\hbar^2/\textrm{MeV}$), and Coriolis attenuation factors $\xi$ used in the PRM calculations are listed in Table~\ref{tab1}.  For each configuration, a normalization of PRM energy of the band-head of each band to the experimental values has been done. However, for the chiral doublet 
bands, only the band-head of the yrast band has been normalized because the energy difference between the doublet bands is given by the model.

\begin{table*}

  \caption{The unpaired nucleon configurations, quadrupole deformation parameters $(\beta, \gamma)$ obtained by CDFT as well as the quadrupole deformation parameters $(\beta^\prime,\gamma^\prime)$, moments of inertia $\mathcal{J}_0$ (unit $\hbar^2/\textrm{MeV}$), and Coriolis attenuation factors $\xi$ used in the PRM calculations for bands D1-D6.}\label{tab1}
\begin{center}
 \begin{tabular}{ccccccc}
 \hline
 \hline
Band & $\rm ~~Unpaired~nucleons~~$ & $(\beta, \gamma)$ & $(\beta^\prime,\gamma^\prime)$ & $\mathcal{J}_0$ & $\xi$\\
\hline
D1 & $\nu(1h_{11/2})^{-1}$ & $~(0.22, 32.8^\circ)~$ &  $~(0.22, 32.8^\circ)~$  & $~16.0~$ & $~0.90~$ \\
D2, D3 & $~~\nu(1h_{11/2})^{-1}\otimes \pi(1h_{11/2})^1 \pi(1g_{7/2})^{-1}~~$ & $(0.20, 28.9^\circ)$ & $(0.20, 20.9^\circ)$ & 26.0 & 0.94 \\
D4-low & $\nu(1h_{11/2})^{-3}$ & $(0.19, 32.7^\circ)$ & $(0.19, 50.0^\circ)$ & 26.0 & 1.00 \\
D4-high & $\nu(1h_{11/2})^{-3}\otimes \pi(2d_{5/2})^2$ & $(0.18, 52.2^\circ)$ & $(0.18, 52.2^\circ)$ & 33.0 & 1.00 \\
D5, D6 & $\nu(1h_{11/2})^{-1}\otimes \pi(1h_{11/2})^2$ & $(0.21, 29.5^\circ)$ & $(0.21, 23.5^\circ)$ & 27.5 & 1.00 \\
D5-high & $\nu(1h_{11/2})^{-1}\otimes \pi(1h_{11/2})^2 \pi(2d_{5/2})^2$ & $(0.20, 32.3^\circ)$ & $(0.20, 32.3^\circ)$ & 33.0 & 0.91 \\
\hline
\end{tabular}
\end{center}
\end{table*}

\begin{figure}[!]
  \vskip-.cm
    \centering
 \includegraphics[width=8.5cm]{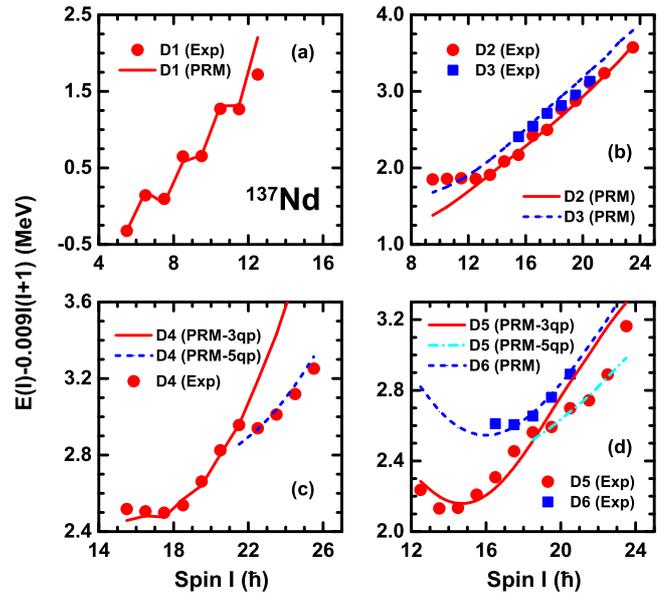}
     \caption{(Color online) Comparison between the experimental energies relative to a rigid rotor (symbols) and the particle-rotor calculations (lines) for the bands discussed in the present work.}
  \label{fig5}
\end{figure}

The calculated excitation energies relative to a rigid rotor are shown in Fig. \ref{fig5}. As one can see in Fig. \ref{fig5}(a), a good description of band D1 built on the $\nu h_{11/2}[505] 11/2^-$ Nilsson orbital is obtained.
The excitation energies of bands D2 and D3, as well as those of the bands D5 and D6, are very similar. These four bands correspond, one to one, to the bands identified in $^{135}$Nd \cite{135nd-lv}. However, the agreement between the theoretical curves and the data points is poor at lower spins, in particular for band D2, due to strong mixing with many other states present in the same energy region, not all shown in Fig. \ref{fig1} but reported previously in Ref.\cite{137nd-npa}, which is beyond the PRM calculation.

\begin{figure}[!]
  \vskip-.cm
  \centering
 \includegraphics[width=8.5cm]{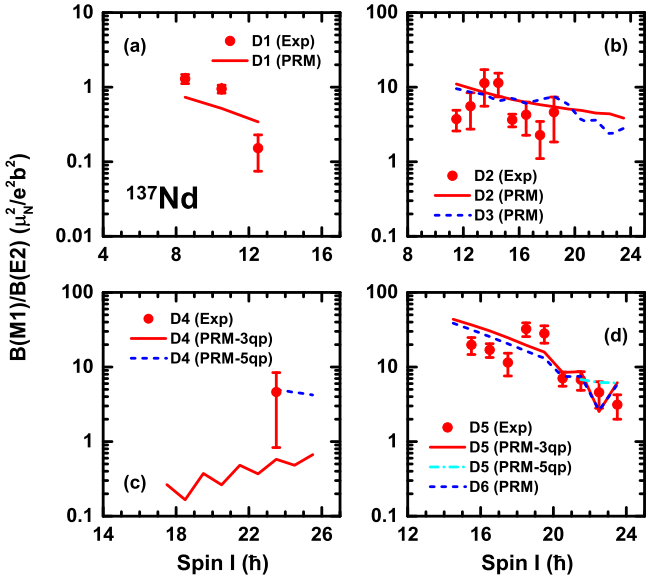}
      \caption{(Color online) Comparison between experimental ratios of transitions probabilities B(M1)/B(E2) (symbols) and the particle-rotor calculations (lines). As there are no experimental values for bands D3 and D6, only the calculated ones are drawn in panels (b) and (d), respectively. }
  \label{fig6}
\end{figure}

The previous interpretation of the three-quasiparticle negative-parity bands in $^{137}$Nd  \cite{137nd-npa}, as well as the  proposed chiral interpretation of the same bands \cite{137nd-brant}, are revised. We  invert the configurations assigned to bands D4 and D5: the new configuration assigned to band D4 below the crossing at $I=41/2\hbar$ is $\nu h_{11/2}^{-3}$, while that of band D5 below the crossing at $I=37/2\hbar$ is $\nu h_{11/2}^{-1}\otimes \pi h_{11/2}^2$. These revised configuration assignments better account for the $B(M1)/B(E2)$ values of the bands as shown in Fig. \ref{fig6}, where the experimental values are determined by assuming zero mixing ratios of the dipole transitions, assumption based on the DCO (Direction Correlations from Oriented states)  published previously in Ref. \cite{137nd-npa}, which are compatible with pure $M1$ character for most dipole transitions.
The same information is obtained from the $R_{ac}$ values deduced in the present work, which are compatible with zero mixing ratios for all M1 transitions.
 The $B(M1)/B(E2)$ values could be extracted for several states of band D5, but only for one state of band D4, because of the too weak and therefore unobserved E2 crossover transitions de-exciting the other states. 
In fact, the observation of $E2$ crossover transitions in band D5 is a sign of significant collectivity, which can be associated with a prolate deformation induced by two protons occupying orbitals at the bottom of the $h_{11/2}$ sub-shell.  The missing $E2$ crossover transitions in band D4 can be due to smaller collectivity of the band, which can be associated to an oblate shape induced by neutrons occupying orbitals at the top of the $h_{11/2}$ sub-shell. On the other hand, the missing $E2$ transitions in bands D3 and D6 are due to the weakness of the bands, which prohibited their observation in the present experiment.

As one can see in Fig. \ref{fig6}, a reasonably good agreement is obtained for the $B(M1)/B(E2)$ values of both bands D2 and D5, giving credit to the assigned configurations. For band D2, the calculated $B(M1)/B(E2)$ values displayed in Fig. \ref{fig6} are obtained employing the $\nu h_{11/2}^{-1}\otimes \pi h_{11/2}^1(g_{7/2})^{-1}$ configuration, which gives a better agreement with the experiment than the configuration with one proton in the $d_{5/2}$ orbital, suggesting therefore the predominance of the proton $g_{7/2}$ orbital in the mixed $\pi (d_{5/2},g_{7/2})$ configuration involved in the bands D2 and D3.

\begin{figure}[]
 \vskip-.cm
\centering
 \includegraphics[width=7.5cm]{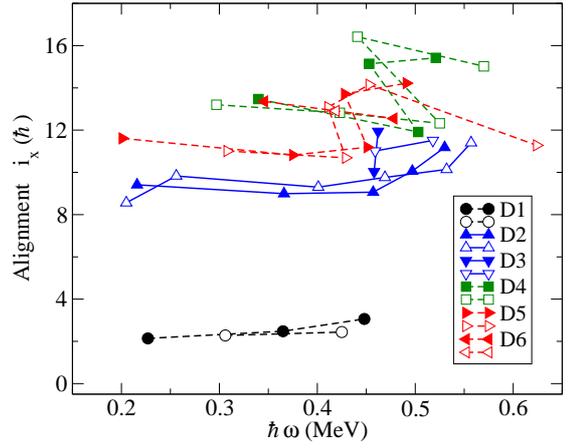}
\caption{(Color online) The experimental quasi-particle alignments for the  chiral rotational bands of $^{137}$Nd.  The Harris parameters used, chosen to result in a flat alignment of band D2, are $\mathcal{J}_0 = 11$ $ \hbar^2$MeV$^{-1}$ and $\mathcal{J}_1=20$ $\hbar^4$MeV$^{-3}$. The two sequences of a given band with even/odd spins are drawn with the same color and with filled/open symbols, respectively. Dashed and continuous lines are used to indicate negative and positive parity, respectively. }
\label{fig7}
\end{figure}

From the single-particle alignments $i_x$ as function of the rotational frequency $\hbar \omega$ shown in Fig. \ref{fig7}, we observe large values of $i_x$ for all bands D2-D5, with a difference of $ 2 \hbar$ between D2 and D3, as well as between D5 and D6, due to either different Fermi levels for protons and neutrons implying the occupation of different Nilsson orbitals, and/or different moments of inertia induced by the active quasiparticles. However, this $2 \hbar$ difference between the doublet bands can be interpreted as possibly resulting from the presence of chiral vibration, like in the case of the $^{135}$Nd nucleus (see e. g.  \cite{135nd-zhu,135nd-lv}).

The $\approx 7 \hbar$ alignment difference between band D2 and band D1 based the one-quasiparticle $\nu h_{11/2}^{-1}$ configuration, clearly indicates a 3-quasiparticle configuration for band D2,  with two more nucleons placed in the opposite-parity proton orbitals $h_{11/2}$ and $(d_{5/2},g_{7/2})$, leading to the $\nu h_{11/2}^{-1}\otimes \pi [h_{11/2}^1(d_{5/2},g_{7/2})^{-1}]$ configuration. The $i_x$ values of $ \approx 11 \hbar$ and $ \approx 13 \hbar$ of the negative-parity bands D5 and D6, respectively, are larger than those of the positive-parity bands D2 and D3, clearly indicating the presence in their configurations of two more protons in the $\pi h_{11/2}$ orbital, leading to the $\nu h_{11/2}^{-1}\otimes \pi h_{11/2}^2$ configuration.

A sudden increase of the single-particle alignment of $ 2\hbar$ is observed in both bands D4 and D5 at slightly different rotational frequencies, which was investigated by calculating different possible additional aligned quasiparticles. A similar crossing could exist in band D6, but could be sharper and not observed in the present experiment in which the weak band D5 has been observed up to lower spin than band D5. It appears that the alignment of two protons in the  $\pi d_{5/2}$ orbital well reproduces both the excitation energies and the $B(M1)/B(E2)$ ratios for both bands D4 and D5, as shown in Figs. \ref{fig4} and \ref{fig6}. We therefore assign  $\nu h_{11/2}^{-3} \otimes \pi d_{5/2}^2$ and  $\nu h_{11/2}^{-1}\otimes \pi h_{11/2}^2 \otimes \pi d_{5/2}^2$ configurations to bands D4 and D5 above the crossing, respectively.

 \begin{figure}[!ht]
  \vskip-.cm
  \centering
 \includegraphics[width=8.5cm]{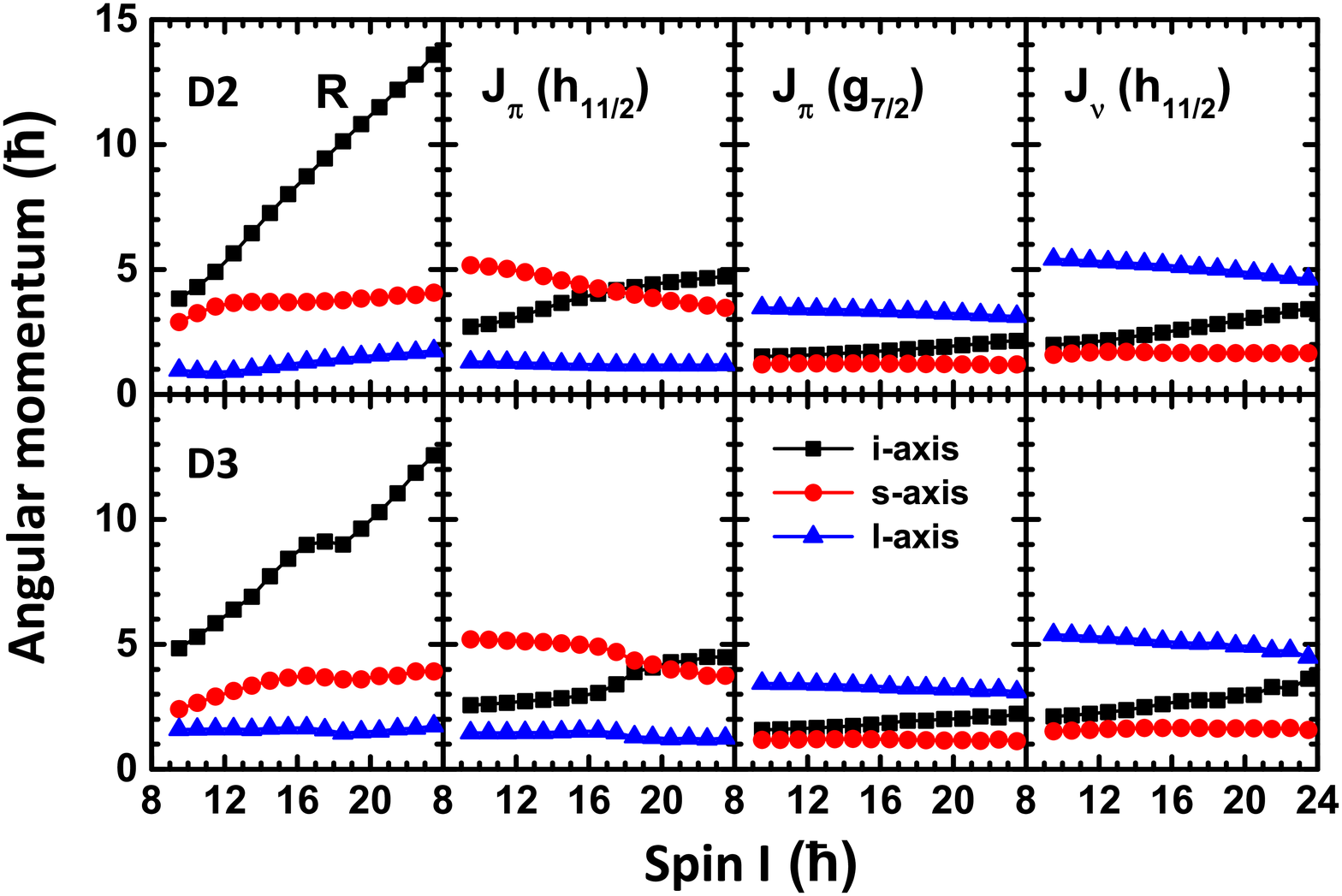}
 \includegraphics[width=8.5cm]{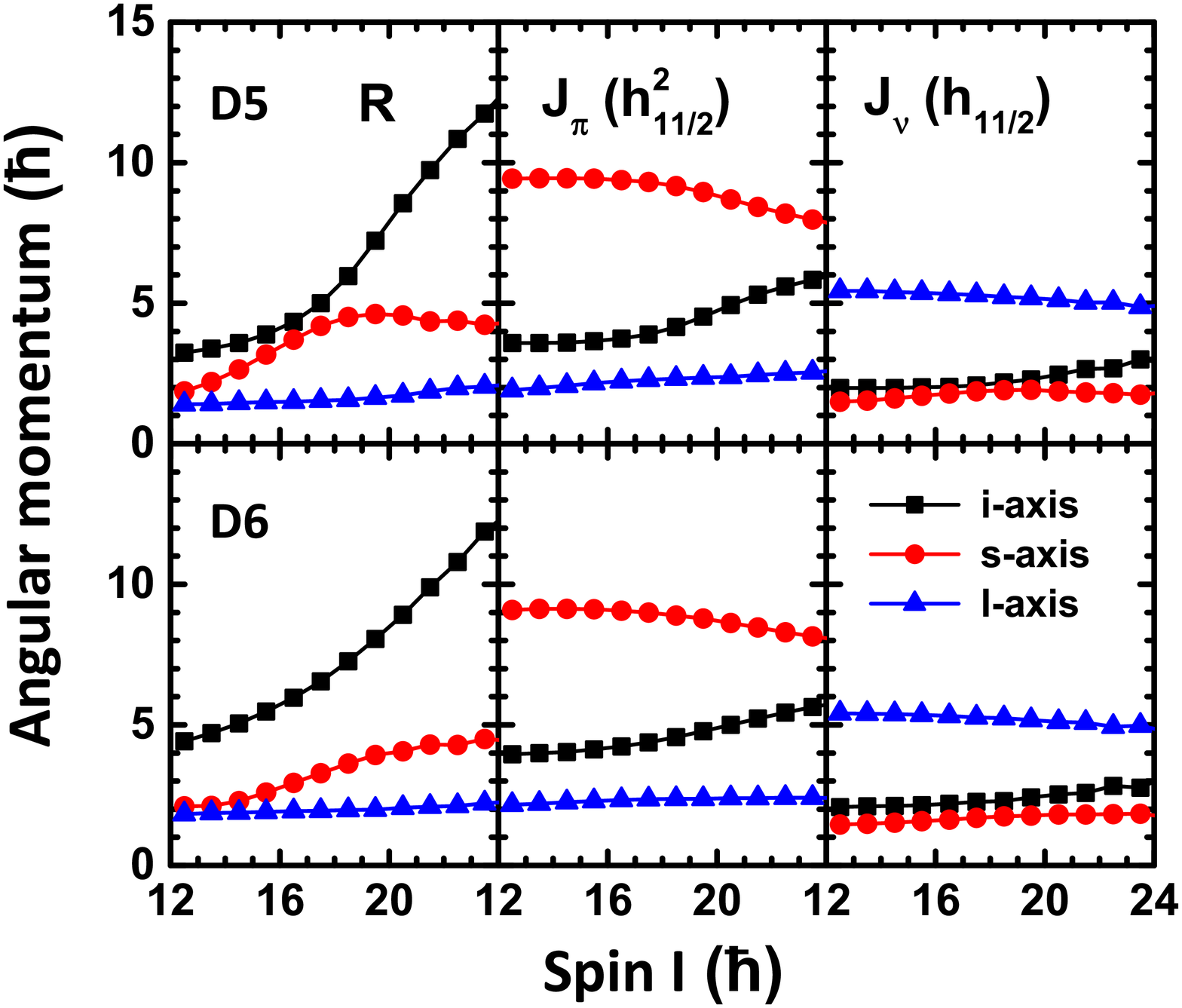}
      \caption{(Color online) The root mean square components along the intermediate (i-, squares), short (s-, circles) and long (l-, triangles) axes of the rotor $R$, valence protons $J_{\pi}$, and valence neutrons $J_{\nu}$ angular momenta calculated as functions of spin by PRM for the chiral partners D2, D3 and D5, D6. One should note that there are three panels for the configuration assigned to bands D2 and D3, which involves three different orbitals $\pi h_{11/2}$, $\pi g_{7/2}$ and $\nu h_{11/2}$, and only two panels for the configuration assigned to bands D5 and D6, which involves two protons and one neutron placed in the $\pi h_{11/2}$ and $\nu h_{11/2}$ orbitals, respectively. }
  \label{fig8}
\end{figure}

The configuration $\nu h_{11/2}^{-1} \otimes \pi [h_{11/2}^1(d_{5/2},g_{7/2})^{-1}]$ assigned to band D2, is similar to that assigned to the corresponding bands of $^{135}$Nd \cite{135nd-lv} and $^{133}$Ce \cite{Ayangeakaa2013Phys.Rev.Lett.172504}. The  alignment of the new band D3, larger by $2\hbar$ than that of band D2, is different from the nearly identical alignments of the corresponding positive-parity bands in $^{135}$Nd and $^{133}$Ce. On the other hand, the same $ 2\hbar$ difference in alignment is present in the negative-parity chiral doublet D5 and D6 of $^{135}$Nd \cite{135nd-lv}, which was interpreted as chiral vibration \cite{135nd-zhu}.  
A similar difference of $2\hbar$ between the $i_x$ values of bands D5 and D6 is also observed. 
This strongly supports the interpretation in terms of chiral vibrations of both doublet bands (D2, D3) and (D5, D6) of $^{137}$Nd.

Finally, the composition of the angular moments in the different assigned configurations has been investigated, in particular for the two pairs of doublet bands (D2, D3) and (D5, D6), which are shown in Fig. \ref{fig8}. One first observes a similar composition of the angular momentum in the two partners of each doublet, which is shared between the three axes of the triaxial core, in agreement with the chiral interpretation. For the (D2, D3) doublet bands the component along the long axis is larger, due to the summed contribution of the high-$\Omega$ $\pi g_{7/2} [504]7/2^+$ and $\nu h_{11/2} [505]11/2^-$ orbitals, while for the (D5, D6) doublet bands the component along the short axis is larger, due to the summed contribution of two protons in the low-$\Omega$ $\pi h_{11/2}[550]1/2^-$ orbital.

\section{Summary}
In summary, we have identified two new bands in $^{137}$Nd, which are interpreted as chiral partners of two previously known three-quasiparticle bands involving one $h_{11/2}$ neutron and two protons placed either in opposite-parity or identical-parity orbitals. Two pairs of chiral bands are therefore present in $^{137}$Nd, like in the case of $^{135}$Nd and $^{133}$Ce, bringing support to the M$\chi$D phenomenon currently explored in the $A=130$ mass region. The difference in the aligned single-particle spin between the partners is interpreted as due to chiral vibration. The configuration of the doublet bands are investigated theoretically using the constrained CDFT and PRM, which reveals the chiral geometry in both chiral doublets. The present results extend the limits of the region of M$\chi$D phenomenon in the $A=130$ mass region, which includes now four Nd nuclei, from $^{135}$Nd to $^{138}$Nd.  The present results encourage further investigations of possible candidates for multiple chiral bands in other isotopes from this and other mass regions.

\section{Acknowledgements}
This work has been supported by the Academy of Finland under the Finnish Centre of Excellence Programme (2012-2017); by the EU 7th Framework Programme Project
No. 262010 (ENSAR); by the National Key R\&D Program of China (Contract No. 2018YFA0404400 and No. 2017YFE0116700), by the National Natural Science Foundation of China (Grants No. 11621131001 and No. 11875075); by the GINOP-2.3.3-15-2016-00034, National Research, Development and Innovation Office NKFIH, Contracts No. PD 124717 and K128947; by the Polish National Science Centre (NCN) Grant No. 2013/10/M/ST2/00427; by
the Swedish Research Council under Grant No. 621-2014-5558; and by the National Natural Science Foundation of
China (Grants No. 11505242, No. 11305220, No. U1732139, No. 11775274, and No. 11575255), and by the National Sciences and Engineering Research of Canada. The use of germanium detectors from the GAMMAPOOL is acknowledged. The work of Q.B.C. is supported by Deutsche Forschungsgemeinschaft (DFG) and National Natural Science Foundation of China (NSFC) through funds provided to the Sino-German CRC 110 ``Symmetries and the Emergence of Structure in QCD" (DFG Grant No. TRR110
and NSFC Grant No. 11621131001). I.K. was supported by National Research, Development and Innovation Office NKFIH, 
  contract number PD 124717. The authors are indebted to M. Loriggiola for his help in target preparation.

\end{document}